\documentclass{article}
\usepackage{graphicx} 
\usepackage{enumerate} 
\usepackage[square]{natbib}
\usepackage{amssymb} 
\usepackage{amsthm} 

\newtheorem{mydef}{Definition} 

\usepackage{authblk} 
\date{\today}

\begin{document}

\title{Einstein's 1935 papers: EPR=ER?}
\author[1]{Gerd Ch.Krizek}
\affil{University of Vienna - Faculty of physics - Quantum particle Workgroup}
\affil{UAS Technikum Wien - Department of Applied Mathematics and Sciences}


\maketitle

\begin{abstract}

\noindent 
In May of 1935, Einstein published with two co-authors the famous EPR-paper about entangled particles, which questioned the completeness of Quantum Mechanics by means of a gedankenexperiment. Only one month later, he published a work that seems unconnected to the EPR-paper at first, the so called Einstein-Rosen-paper, that presented a solution of the field equations for particles in the framework of general relativity. Both papers ask for the conception of completeness in a theory and, from a modern perspective, it is easy to believe that there is a connection between these topics. We question whether Einstein might have considered that a correlation between nonlocal features of Quantum Mechanics and the Einstein-Rosen bridge can be used to explain entanglement. We analyse this question by discussing the used conceptions of "completeness," "atomistic structure of matter," and "quantum phenomena." We discuss the historical embedding of the two works and the context to modern research. Recent approaches are presented that formulate an EPR=ER principle and claim an equivalence of the basic principles of these two papers.


\end{abstract}

\section{Introduction}
\parindent 0cm       

The historical approach to the foundations of physics is often misunderstood as a historical work alone. To study specific questions in the history of physics is a method of introspection about problems that are still relevant to recent physics. This work is not intended to be a historical work alone. Recent aspects of different conceptions of completeness will be discussed, starting from a historical perspective. The connection between these two papers (which at first look seem to be independent) is presented and provides important insights into Einstein's conception of reality and the recent advent of the EPR=ER concept.     \newline 

The year 1935 is a remarkable year in the context of Quantum Mechanics. The discussion on the incompleteness of Quantum Mechanics found its culmination in the famous EPR paper \cite[]{einstein1935can} and the reply by \cite[]{bohr1935can}. Further, \cite[]{schrodinger1935gegenwartige} expressed his criticism on the oddities of Quantum Mechanics with his famous Gedankenexperiment of a cat in a box. 

At first view unconnected to the discussion in Quantum Mechanics, \cite[]{einstein1935particle} published one month later an article which is famous as well, but in a completely different field of physics. It is the paper that presented the Einstein-Rosen bridge solution in General Relativity, which later became well-known as ``wormhole'' in popular science and science fiction.

We will show that the EPR paper cannot be seen as stand-alone, but must be seen in context with the \cite[]{einstein1934method} paper and the \cite[]{einstein1935particle} paper. 


In 1934, Einstein expressed his point of view on physical theories in general, the formation of new theories, the connection to an ontology, and the concept of reality. Later articles and letters from Einstein will help to complete the picture about his views on the nature of reality and his notion of completeness. 

It is tempting to think that the problem Einstein illustrated with the EPR paper is meant to be solved with the Einstein-Rosen bridge solution by solving the occurring nonlocality\footnote{It is still a matter of dispute in the interpretations of Quantum Mechanics if there is nonlocality in Quantum Mechanics. We define a nonlocal theory as a theory that violates the Bell inequalities.} in Quantum Mechanics with a two-sheet wormhole that represents the entangled particles. After analysing the EPR paper and the notions of reality and completeness, the idea of a connection between those papers will be discussed in context with recent developments in the field. 

\section{Einstein's view on theories and theory formation}
\label{section view on theories}
In 1934 Einstein was driven by the long-lasting debates on the foundations of Quantum Mechanics to clarify his views on the method of theoretical physics in general. He emphasized the nature of the constituents of a theory used in physics \cite[]{einstein1934method}:

\begin{quote}
\textit{``For to the discoverer in that field, the constructions of his imagination appear so necessary and so natural that he is apt to treat them not as the creations of his thoughts but as given realities.''}
\end{quote}

and

\begin{quote}
\textit{``He is in just the same plight as the historian, who also, even
though unconsciously, disposes events of the past around ideals
that he has formed about human society.''}
\end{quote}

This already indicates that the objection that Einstein was a proponent of a naive realism\footnote{See\cite[page 309ff]{janssen2014cambridge}} has to be reconsidered. His ontology is, of course, not at trivial one, and we have to define what is meant exactly by naive realism. We will come back later to that discussion. 

Einstein then mentions the second elementary truth science is based upon; its dependency on experience, the need for empirical verification: 

\begin{quote}
\textit{``Pure logical thinking can give us no knowledge whatsoever of the world of experience; all knowledge about reality begins with experience
and terminates in it.''}
\end{quote}

It is remarkable that Einstein proposed his views parallel to \cite[]{popper2005logic}, and they agree on several points. We will see later 
that their views are connected more deeply. What are those ideals that Einstein is referring to? In his view, a theory is built upon concepts and laws and has to obey certain requirements:




\begin{quote}
\textit{``The basic concepts and laws which are not logically further reducible constitute the indispensable and not rationally deducible part of the theory. It can scarcely be denied that the supreme goal of all theory is to make the irreducible basic elements as simple and as few as possible without having to surrender the adequate representation of a single datum of experience''}
\end{quote}
Einstein is referring to Ockham\textsc{\char13}s razor principle in this statement on the economy that the theory building process has to comply with. A detailed analysis of Ockham\textsc{\char13}s razor in context to Quantum Mechanics is given by \cite[]{krizek2017ockham}. \newline

Einstein's ideas on theory building are guided by the structure of the general Theory of Relativity and its success. To Einstein, it was representative to show the incorrectness of the induction principle as a theory building process and, to the contrary, consolidated his beliefs in a more top-down-orientated theory building process out of principles. 


By rejecting induction as theory building principle and taking up the position that the elements of theories are creations of human thoughts, the question arises if there is a correct way to find insights about nature. And here Einstein makes clear that to him there is such a way \cite[]{einstein1934method}:

\begin{quote}
\textit{`` To this I answer with complete assurance, that in my opinion there is the correct path and, moreover, that it is in our power to find it. Our experience up to date justifies us in feeling sure that in Nature is actualized the ideal of mathematical simplicity. It is my conviction that pure mathematical construction enables us to discover the concepts and the laws connecting them which give us the key to the understanding of the phenomena of Nature.''}
\end{quote}

His argument in favour of this position is the huge success of mathematics as a tool to describe our experiences about nature. It is this unreasonable effectiveness of mathematics that led Eugene \citeauthor{wigner1960unreasonable} to publish his paper \cite[]{wigner1960unreasonable} of the same title. On the other hand, Einstein identified a problem in the above argument, when he stated that this mentioned success is claimed as well by theories that cannot deliver a deeper ontological insight, such as classical mechanics.

\begin{quote}
\textit{``Have we any right to hope that experience will guide us aright, when there are theories (like classical mechanics) which agree with experience to a very great extent, even without comprehending the subject in its depths?''}
\end{quote}

Einstein ignores this objection and heads for a worldview with a distinguished position of mathematics:\footnote{Wigner will refer to Einstein\textsc{\char13}s objection 25 years later.} 


\begin{quote}
\textit{``...the truly creative principle resides in mathematics. In a certain sense, therefore, I hold it to be true that pure thought is competent to comprehend the real, as the ancients dreamed.''}
\end{quote}

It reflects a kind of mathematical realism, which is in one aspect contradictory, when Einstein pointed out that the elements of the theory are ideals, are pure constructions of human thoughts, but on the other hand represent a ``correct way,'' reflecting a truth about nature. 



As a basic entity of this worldview, he adopts the four dimensional continuum, which already was extremely successful for unification of physics during the end of the 19th century. He refers to this success and the principle of simplicity and states:

\begin{quote}
\textit{``It is essential for our point of view that we can arrive at these constructions and the laws relating them one with another by adhering to the principle of searching for the mathematically simplest concepts and their connections. In the paucity of the mathematically existent simple field-types and of the relations between them, lies the justification for the theorist's hope that he may comprehend reality in its depths.''}
\end{quote}

Einstein presented and clarified his ideas on theory formation in general, about his ontological ideas and the scientific method in general. He presented field theories as candidates for this simple new theory that describes the real. His concern was mostly about how to integrate the atomistic structure\footnote{In today's notion, we would call it quantum nature} of matter and energy. 

\newpage

We will see later that he used this wording again in his 1935 papers, and it is relevant to see what he meant exactly: 

\begin{quote}
\textit{``...the theory\footnote{He refers to the field theory} in its basic principles is not an atomic one in so far as it operates exclusively with continuous functions of space, in contrast to classical mechanics whose most important feature, the material point, squares with the atomistic structure of matter.''}
\end{quote}

What Einstein understands as ``atomic theory'' or ``atomistic structure'' is discrete localization in space, in contrast to continuous functions of fields in space. And he emphasizes the importance of this idea in the conclusion of the paper:

\begin{quote}
\textit{``Thus in a continuum theory, the atomistic character could be satisfactorily expressed by integral propositions without localizing the particles which constitute the atomistic system. Only if this sort of representation of the atomistic structure be obtained could I regard the quantum problem within the framework of a continuum theory as solved.''}
\end{quote}

To Einstein, this concept is of great importance to represent particles in a field theory. The singularity-free localization of particles is a core element of a successful theory. Some aspects of Einstein's views on the structure of theoretical physics directly relate to the nature of reality. We will now analyse the notions of Einstein's reality and the structure of theories.


\newpage

\section{Einstein`s not-naive realism and the notions of reality}

\subsection{A classification scheme for physical theories}
\label{section Classification scheme}

A classification scheme is presented to categorize different aspects of a theory. Therefore, we define graphical representations of physical quantities, physical laws, physical concepts and elements of reality.

\begin{table}[h]
\centering
\label{SchemeTheory}
\begin{tabular}{cc}

\begin{tabular}[c]{@{}c@{}}\\ \includegraphics[width=0.05\textwidth, height=30px]{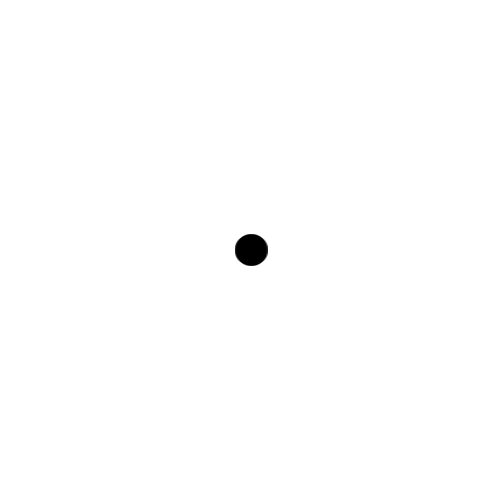} \\ \\ \end{tabular}  & \begin{tabular}[c]{@{}c@{}}  \\ Physical quantity \\ \end{tabular} \\ 
\begin{tabular}[c]{@{}c@{}}\\ \includegraphics[width=0.2\textwidth, height=50px]{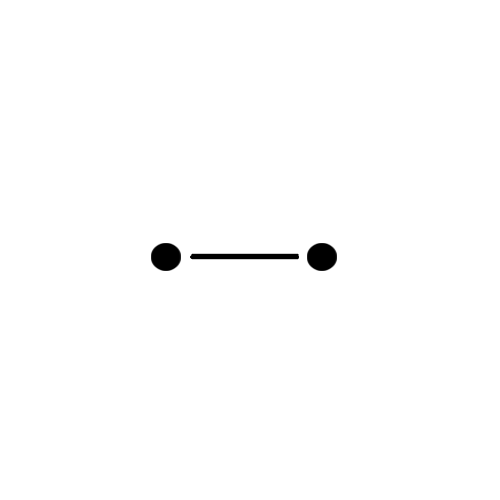} \\ \\ \end{tabular} & \begin{tabular}[c]{@{}c@{}}  \\ Law between physical quantities \\ \end{tabular} \\ 
\begin{tabular}[c]{@{}c@{}}\\\includegraphics[width=0.3\textwidth, height=80px]{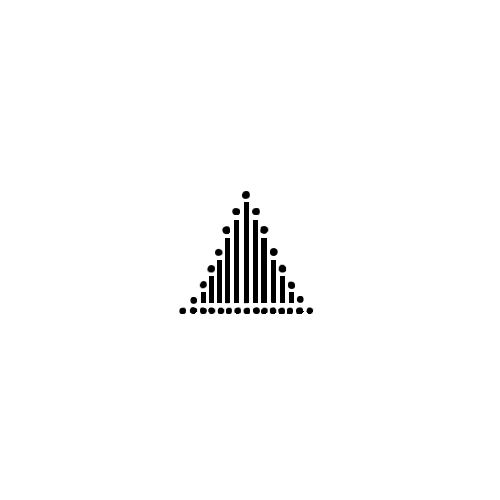} \\ \\ \end{tabular} & \begin{tabular}[c]{@{}c@{}} \\ Concept connecting laws and physical quantities \\ \end{tabular}                                                                                 \\ 
\begin{tabular}[c]{@{}c@{}}\\ \includegraphics[width=0.3\textwidth, height=90px]{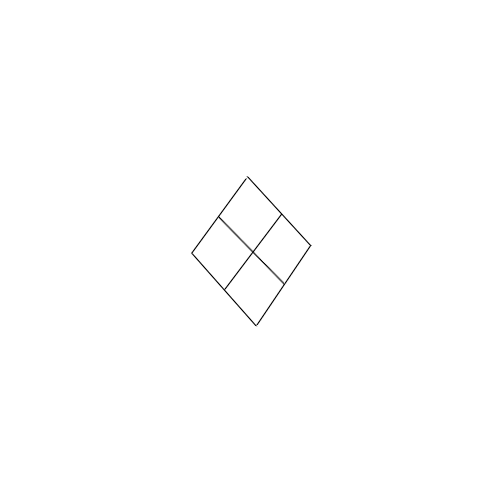} \\ \\ \end{tabular} & \begin{tabular}[c]{@{}c@{}}\\ Elements of objective reality \\ \end{tabular}                      \\ 
\end{tabular}
\caption{Icons used in the classification scheme}
\end{table}

Out of these icons, we can construct a depiction for a theory and for elements of reality inside objective reality.

\begin{figure}[h]
\centering

\includegraphics[width=0.4\textwidth]{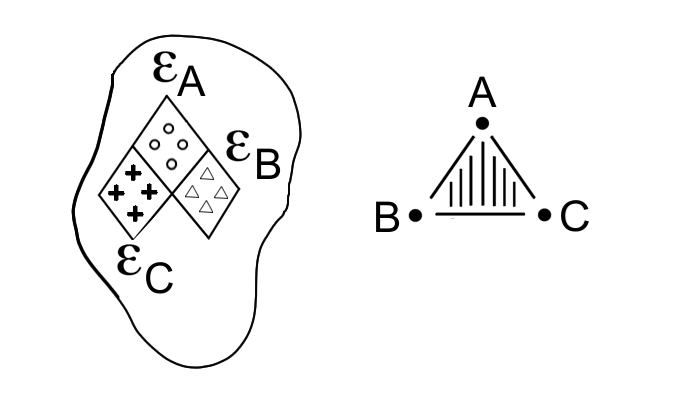}

\caption{Elements of reality and the structure of a theory}
\end{figure}

\newpage
Formally, we define a set of elements of a theory $ T_{PR} $ and a set of elements of reality $ PR $.

\begin{equation}
T_{PR} = \{ t_i \} = \{ A,B,C \}
\label{Definition Elements of theory}
\end{equation}
\begin{equation}
PR = \{\epsilon_i\} = \{ \epsilon_A,\epsilon_B, \epsilon_C \}
\label{Definition Elements of reality}
\end{equation}

We will make use of them in the analysis of the reality criterion and the conceptions of completeness. \newline

For classification, we propose four levels in a theory to classify physical theories regarding their different aspects. The intention is to compare different interpretations of Quantum Mechanics regarding their formalism, concepts and ontologies. 

\begin{itemize}

\item The first level contains the mathematical formalism of the theory, mathematical symbols and the connecting laws. The mathematical symbols can be identified with symbols for physical quantities, so we will speak of physical quantities, although we mean only their symbolic representations. 
\item The second level contains the assignment of the previously defined physical quantities to sensations. Physical quantities consist of quantities and units, and need a procedure to be measured. All these assignments are contained in the second level.
\item The third level contains concepts and principles that give reason and structure to the theory. First principles, such as extremal principles or conservation principles, belong to this level. 
\item The fourth level is the ontological embedding of the theory; it defines how physical quantities and concepts are related to ontological elements. Even if there are no ontological elements, such as in a positivistic philosophy of science, this would be defined here. Here, the elements of reality Einstein refers to are located. They are parts of the objective reality.

\end{itemize}

We will use the classification scheme for the discussion of the concept of completeness, the criterion of reality, and in the analysis of naive realism in the chapter \ref{section Naive realism}. A detailed proposal of the classification scheme can found in \cite{krizek2017classification}.

\subsection{Naive realism}
\label{section Naive realism}
Naive realism, or direct realism, is the view that the perception delivers a full mapping between the elements of reality and the sensations. There are a lot of subtleties that could be discussed in that regard; some views of naive realism will require the spatiotemporal reference frame, some views will assume materialistic ideas. We will confine ourselves to those aspects that are of interest for our discussion.\newline

We already mentioned the objection that Einstein was a proponent of naive realism. The term "naive" is depreciative inherently and includes prejudices, though will we make use of it due to the historical account, but we distance ourselves from a depreciative intent of this term. To get a better idea about the concept of naive realism, we present the definition by \cite[]{russell1940inquiry}: 

\begin{quote}
\textit{``We all start from naive realism, i.e., the doctrine that things are what they seem. We think that grass is green, that stones are hard, and that snow is cold. But physics assures us that the greenness of grass, the hardness of stones, and the coldness of snow are not the greenness, hardness, and coldness that we know in our own experience, but something very different. The observer, when he seems to himself to be observing a stone, is really, if physics is to be believed, observing the effects of the stone upon himself.''}
\end{quote}

Another statement of naive realism by \cite[]{einstein1946remarks} shows Einstein's position on this question: 

\begin{quote}
\textit{``...naive realism,... according to which things “are” as they are perceived by us through the senses. This illusion dominates the daily life of men and of animals; it is also the point of departure in all of the sciences, especially of the natural sciences. ... The overcoming of naive realism has been relatively simple.''}
\end{quote}

\begin{figure}[h]
\centering

\includegraphics[width=0.7\textwidth]{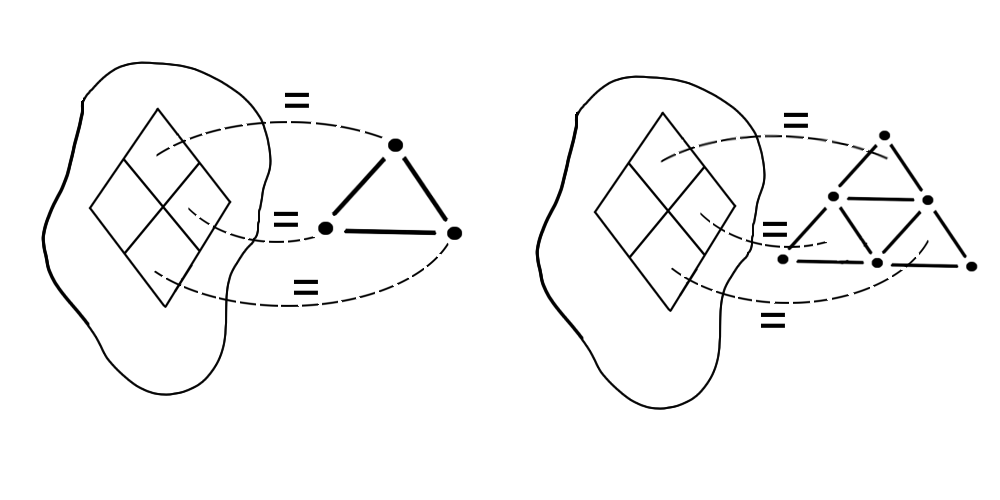}

\caption{Naive realism and an ontology for quantities and concepts}
\end{figure}

This last remark in \cite[]{einstein1946remarks} shows that Einstein would not consider himself as a proponent of naive realism. 

If we apply the classification scheme presented before on a definition of naive realism, one would conclude that there are identities between the physical quantities, or the physical concepts with the ontological elements of reality. \newline

In the context of materialism, this would mean that the rigid bodies we experience through sensations and we model as a concept within Newtonian mechanics would be assumed to exist as experienced. This view is rejected by physics and philosophy of science. A detailed discussion on the rejection of direct and naive realism can be found in \cite[]{morvan2004arguments}. \newline

But still, a kind of naive realism is discussed in recent physics, the mathematical universe hypothesis(MUH) \cite[]{tegmark2004parallel}:

\begin{quote}
\textit{``Let us now digest the idea that physical world (specifically, the Level III multiverse) is a mathematical structure. Although traditionally taken for granted by many theoretical physicists, this is a deep and far-reaching notion. It means that mathematical equations describe not merely some limited aspects of the physical world, but all aspects of it. It means that there is some mathematical structure that is what mathematicians call isomorphic (and hence equivalent) to our physical world, with each physical entity having a unique counterpart in the mathematical structure and vice versa.''}
\end{quote}

It is not our goal to discuss the MUH here, but we have seen before that Einstein had in mind a kind of mathematical realism. We will discuss the notions of Einstein's realism later, after clarifying some subtleties in the EPR paper. 

%

\subsection{Subtleties in the EPR paper}

When we discuss reality, we have to clarify the used terms, and there we immediately face one subtlety. In the handwritten note for the talk he held in Oxford at the Herbert Spencer lecture on the method of theoretical physics, \cite[]{einstein1934zurMethodeHandschriftlich} made use of the two synonymical terms ``Wirklichkeit'' and ``Realit\"at''. In the English translation and published paper \cite[]{einstein1934method} both terms merge to the term ``reality.'' It is clear from the context in which Einstein made use of these terms that ``Wirklichkeit'' stands for the objectively given facts in nature, independent of observation. ``Wirklichkeit'' refers to the ontology in the world.  He makes use of ``Realit\"at''only once in the context of an allegory to explain his personal struggle with philosophy of science \cite[]{einstein1934method}:     

\begin{quote}
\textit{`` For to the discoverer in that field\footnote{Theoretical physics}, the constructions of his imagination appear so necessary and so natural that he is apt to treat them not as the creations of his thoughts but as given realities.''}
\end{quote} 
\newpage

In \cite[]{einstein1935can}, we find three wordings, but the paper was written originally in English: ``objective reality,'' ``physical reality'' and ``reality.'' The latter is just used as a short form for ``physical reality''. The definition of ``objective reality'' can be found in the introduction of the paper:

\begin{quote}
\textit{``Any serious consideration of a physical theory must take into account the distinction between the objective reality, which is independent of any theory, and the physical concepts with which the theory operates.''}
\end{quote}

It is noticeable that Einstein defines the term ``objective reality'' and then in the definition of completeness makes use of ``physical reality'' (see citation in Section \ref{SectionCompleteness}), which has not been defined before \cite[]{einstein1935can}:

\begin{quote}
\textit{``Whatever the meaning assigned to the term complete, the following requirement for a complete theory seems to be a necessary one: every element of the physical reality must have a counterpart in the physical theory''}
\end{quote}

Out of how the reality criterion refers to this ``physical reality'' and the elements of reality, and the lack of further definition of ``physical reality'' we conclude that ``physical reality'' is synonymous with ``objective reality''. \newline

Out of Einstein's dissatisfaction with the precipitate submission of the manuscript by Boris Podolsky \cite[letter 206, p.537]{vonMeyenn2010entdeckung}, we conjecture that some flaws remained in the paper. \newline

One more of these flaws is the sloppy use of the assignment of physical quantities to elements of physical reality, which we demonstrate by means of three statements from the paper.

\subsubsection*{Statement 1}

\begin{quote}
\textit{``... there is an element of physical reality corresponding to the physical quantity A.''}
\end{quote}

\subsubsection*{Statement 2}

\begin{quote}
\textit{``The usual conclusion from this in Quantum Mechanics is that when the momentum
of a particle is known, its coordinate has no physical reality.''}
\end{quote}

\subsubsection*{Statement 3}

\begin{quote}
\textit{``In accordance with our criterion of reality, in the first case we must consider the quantity P as being an element of reality, in the second case the quantity Q is an element of reality.''}
\end{quote}

There is a big difference in the nature of reality in the different statements above. If we apply the classification scheme for the structure of theories given by \cite[]{krizek2017classification}, we see that the latter statements would represent a naive realism where the referred physical quantity would be identical with the ontology of the theory. \newline

%


The first statement that uses the correspondence between the physical quantity and the element of reality is more subtle. There ,the physical quantity is a mathematical object with interpretation and corresponds to an independent ontological element of reality (see Table 3). The element of reality itself is only comprehensible through sensations, though it is expected to exist in the physical reality but never accessible to us in its whole nature. \newline

%

It is coherent with Einstein's views on philosophy of science we presented above, and with the context of the EPR-paper, that he did not favour the point of view presented in statements 2 and 3. 

\subsection{Notions of reality}

\subsubsection*{Elements of reality}
\label{Section Elements of reality}

In the EPR argument and the reality criterion specifically, a term is used to describe and speak about properties of objective reality: "element of reality." It is interesting to trace back the term "element" to a letter from Einstein to the philosopher Moritz Schlick in which he refers to Ernst Mach \cite[]{hentschel1986korrespondenz}:

\begin{quote}
\textit{``Es scheint mir nun, da\ss \, das Wort 'wirklich' in verschiedenem Sinne genommen
wird, je nachdem es von Empfindungen oder von Ereignissen bzw. Thatbest\"anden in physikalischem Sinne ausgesprochen wird. Wenn zwei verschiedene V\"olker unabh\"angig voneinander Physik treiben, werden sie Systeme schaffen, die bez\"uglich der Empfindungen ('Elemente' im
Sinne Machs) gewiss \"ubereinstimmen. Die gedankliche Konstruktion, die die
beiden zur Verknf\"upfung dieser 'Elemente' ersonnen, k\"onnen weitgehend verschieden
sein.'' 
\newline
\newline
``It seems to me that the term "real"\footnote{Refer to the discussion before on the terms "wirklich" und "real".} is used in different meaning, depending on whether used in physical context with sensations or on the other hand events respectively facts. When two nations pursue physics independently from each other, they will create systems that agree concerning the sensations(elements in the meaning of Mach). The construction of ideas both of them developed to connect these elements can differ widely.\footnote{Translation by the author}''}
\end{quote}

Is this identification of the Machian elements mentioned by Einstein in the letter to Morith Schlick and the elements used in the EPR argument justified? In the Machian sense, an element is defined as sensation, as a representation for events \cite[]{mach1922analyse}:

\begin{quote}
\textit{``Die Ansicht, welche sich allm\"ahlich Bahn bricht, daß die Wissenschaft sich auf die \"ubersichtliche Darstellung des Tats\"achlichen zu beschr\"anken habe, f\"uhrt folgerichtig zur Ausscheidung aller m\"u\ss igen, durch die Erfahrung nicht kontrollierbaren Annahmen, vor allem der metaphysischen (im Kantschen Sinne). H\"alt man diesen Gesichtspunkt in dem weitesten, das Physische und Psychische umfassenden Gebiete fest, so ergibt sich als erster und n\"achster Schritt die Auffassung der „Empfindungen" als gemeinsame „Elemente" aller m\"oglichen physischen und psychischen Erlebnisse, die lediglich in der verschiedenen Art der Verbindung dieser Elemente, in deren Abh\"angigkeit voneinander bestehen.''
\newline
\newline
``
The view that wins recognition step by step, that science has to confine to clearly arranged representations of facts, leads to the exclusion of all dispensable, through experience not-controllable assumptions, specifically the metaphysical ones(in the meaning of Kant). Applying this view to physical and psychical fields, as first and next consequence results the view of sensations as common "elements" of all possible physical and psychical experiences. These sensations only exist in their different ways of connection between them and their mutual dependence of each other.\footnote{Translation by the author}  
''}
\end{quote}

and further

\begin{quote}
\textit{``Berkeley\footnote{Mach means the philosopher George Berkley} sieht die „Elemente" als durch etwas au\ss er denselben Liegendes, Unbekanntes (Gott) bedingt an, wof\"ur Kant, um als n\"uchterner Realist zu erscheinen, das „Ding an sich" erfindet, w\"ahrend die hier vertretene Anschauung mit einer Abh\"angigkeit der „Elemente" von einander praktisch und theoretisch das Auskommen zu finden glaubt.''}
\newline
\newline
``
\textit{Berkley sees the "elements" as caused by something unknown(god) beyond them, Kant on the other hand, to appear as a down-to-earth realist, invents "the thing itself", while the here presented approach of mutual depending elements claims to get along for practical and theoretical purposes.\footnote{Translation by the author}''}
\end{quote}

Machs philosophy of science is a positivistic one, an Empirio-criticism \cite[]{Avenarius1905menschliche} and \cite[]{Avenarius1906kritik}. Metaphysical parts of the theory are rejected as non-physical and unnecessary, also with regard to a principle of economy of thought all theories have to comply with. The elements Einstein is referring to are those parts of the theory that are indisputable because they represent the experiences, which are independent of the type of physical theory, in Einstein's relativistic context the invariants of a theory. \newline


It is irony that Einstein refers to the originally positivistic assigned term "elements" to defend his realistic point of view, but he mentions also the Machian elements and gives a clear distinction between Machian elements and elements of reality \cite[]{einstein1935can}:

\begin{quote}
\textit{``In a complete theory there is an element corresponding to each element of reality.''}
\end{quote}

The "element" in the theory is an element in the sense of Mach. The corresponding "element of reality" is not part of the theory; it is part of the ontology, part of the objective reality independent of our perception. \newline

It is strange that the term "element" is never again used by Einstein; not in his explanation of the EPR argument in a letter to Schr\"odinger \cite[Letter 206, p.537]{vonMeyenn2010entdeckung}, nor in the explanation in a letter to Popper\cite[]{popper2005logic}, nor in the reformulation of the EPR argument in \cite[]{einstein1948quantum}. This could give rise to the objection that the term "element" was coined by Podolsky who wrote most of the article as pointed out by\cite[Chapter 2.4]{kiefer2015albert} and in \cite[Letter 206, p.537]{vonMeyenn2010entdeckung}. \newline 

The content of the EPR argument has been formulated after long discussions\footnote{\cite[Letter 206, p.537]{vonMeyenn2010entdeckung}} of Einstein, Podolsky, and Rosen, so it seems feasible that the cornerstones of the paper were clearly defined before by all authors. \newline

Secondly, though Einstein did not use the term in this context again, he mentions the "elements" before in the 1917 letter to Moritz Schlick \cite[]{hentschel1986korrespondenz}: 

\begin{quote}
\textit{``...a gap which widens progressively with the developing unification of the logical structure, that is with the reduction in the number of the logically independent conceptual elements required for the basis of the whole system.''}
\end{quote}

He further makes use of them in \cite[]{einstein1934method}:

\begin{quote}
\textit{``It can scarcely be denied that the supreme goal of all theory is to make the irreducible basic elements as simple and as few as possible...''}
\end{quote}

\newpage

It seems that Einstein had not agreed upon the way the EPR argument was constructed, paired with Podolsky's not-agreed submission of the EPR paper, and a parallel press statement raised Einstein's anger \cite[]{kiefer2015albert}. It is therefore plausible that the elements are indeed a reference by Einstein to the positivistic Machian "elements," amending them with ontological counterparts, the realistic "elements of reality," but Einstein did not want to refer to the EPR-paper and its terminology again.

\subsubsection*{The reality criterion}

The reality criterion is defined at the begin of the EPR paper to start the line of argument and identify physical quantities and concepts that have a counterpart in the objective reality. The problem is to find those quantities and concepts that refer to reality and have a corresponding element in reality. For example, the concept of centre of mass of a body is not useful to describe some fundamental real property of the rigid body. It is merely a concept in the theory that is helpful to construct a plain theory and has some practical and technical advantages. So it is needed to find a way to identify elements in the theory that refer to an element in reality. Einstein admits that in his letter to Schr\"odinger \cite[letter 206, p.537]{vonMeyenn2010entdeckung}:

\begin{quote}
\textit{``Die eigentliche Schwierigkeit liegt darin, da\ss \, die Physik eine Art Metaphysik ist; Physik beschreibt „Wirklichkeit“. Aber wir wissen nicht, was „Wirklichkeit“ ist; wir kennen sie nur durch die physikalische Beschreibung!''}\footnote{Translation by the author}
\end{quote}

\begin{quote}
\textit{``The difficulty is due to that physics is a kind of metaphysics; physics describes "reality". But we do not know what "reality" is; we only know it through the physical description!''}
\end{quote}

The identified physical quantities that refer to the elements of reality are then chosen to show the incompleteness of Quantum Mechanics \cite[]{einstein1935can}:

\begin{quote}
\textit{``If, without in any way disturbing a system, we can predict with certainty (i.e., with probability equal to unity) the value of a physical quantity, then there  exists an element of physical reality corresponding to this physical quantity.''}
\end{quote}

\begin{figure}[h]
\centering

\includegraphics[width=0.7\textwidth]{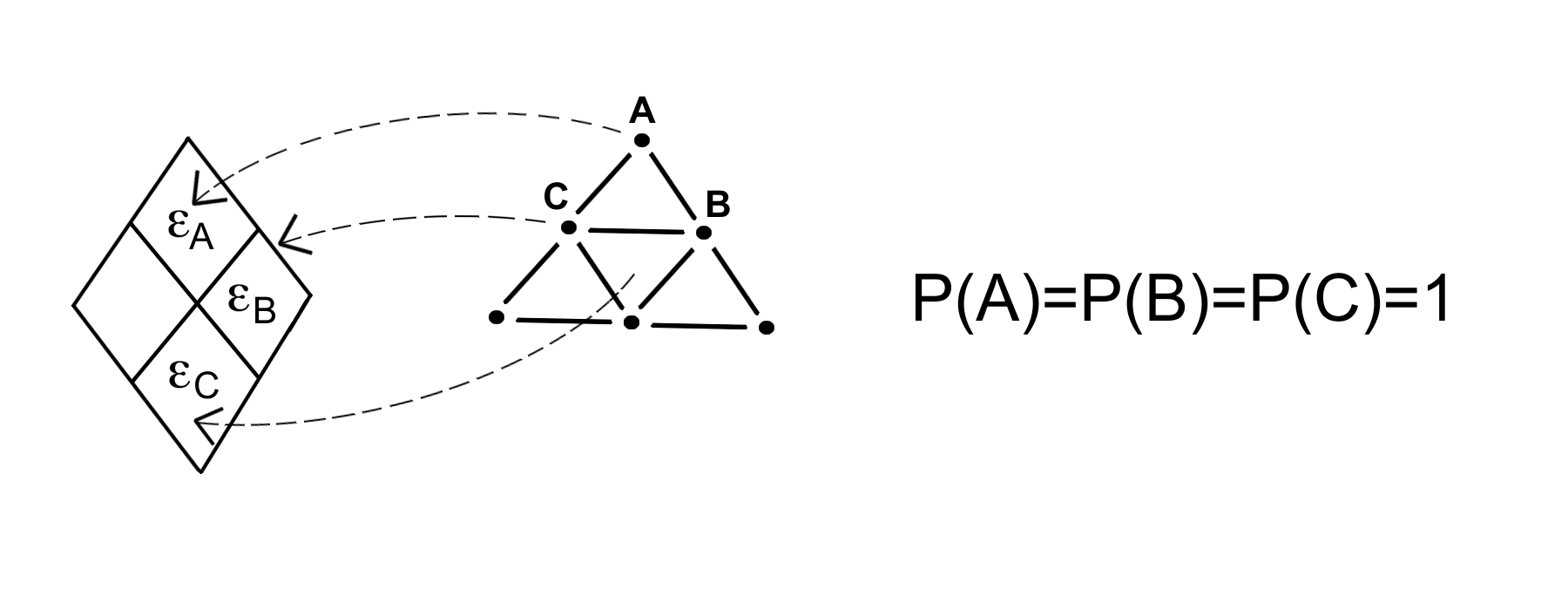}

\caption{Elements of reality}
\end{figure}

We can define the reality criterion more formally by using the definitions from equations \ref{Definition Elements of theory} and \ref{Definition Elements of reality}. \newline

\newpage 

One has to define a probability measure $P(t_i) $ for the elements of the theory. Then the reality criterion can be formally expressed as: 
\begin{equation}
\forall t_i \in T_{PR} \wedge P(t_i)=1: \exists \epsilon_i \in PR
\end{equation}

We will have a mapping $\phi$ that connects the element of the theory with the corresponding element of reality.

\[ \phi\colon T_{PR}\to PR     \] 
\[ \phi(t_A)=\epsilon_A \]

The remarkable thing in the way the reality criterion is defined is the working direction. Out of a probability statement in the theory, the reality criterion identifies corresponding elements of reality. If we assume classical physics that fulfills determinism, all physical quantities obey a probability\footnote{We neglect all interpretational issues concerning probability here and consider probability in classical physics as a propensity.} measure of one. Therefore, in classical physics, all physical quantities should have an corresponding element of reality. The situation is not that trivial, as an example will show. \newline

Let us assume Newtonian mechanics in one dimension to describe the motion; i.e. that is, the position of a particle A in reference to a second particle B. The particle is assumed to have a constant mass and is described by a mass point. Our theory then contains the following elements:  Occurring forces, the mass of the particle, the position coordinate of particle A and the position coordinate of particle B.  

\begin{equation}
T_{PR}  = \{ F,\hspace{5px} m,\hspace{5px} x_A(t),\hspace{5px} x_B(t) \}
\end{equation}

The theory will connect these elements with Newton's second law to describe the motion of particle A

\begin{equation}
F=m\frac{d^2x_A(t)}{dt}
\end{equation}

Let us assume the most simple situation; the force vanishes and becomes zero. This differential equation delivers the equation of motion with two independent constants that refer to the initial conditions. They turn out to be the initial speed of particle A and the initial position of particle A. 

\begin{equation}
x_A(t)=v_{Ao}\cdot t+x_{Ao}
\end{equation}

Our theory is deterministic; therefore, the probability measure is equal to one, so by applying the criterion of reality we can map all elements of the theory to elements of reality with a mapping $\phi$. That way we have identified corresponding elements of reality $PR =  \{ \epsilon_F, \epsilon_m, \epsilon_{x_A(t)}, \epsilon_{x_B(t)} \}$. \newline

\begin{figure}[h]
\centering

\includegraphics[width=0.6\textwidth]{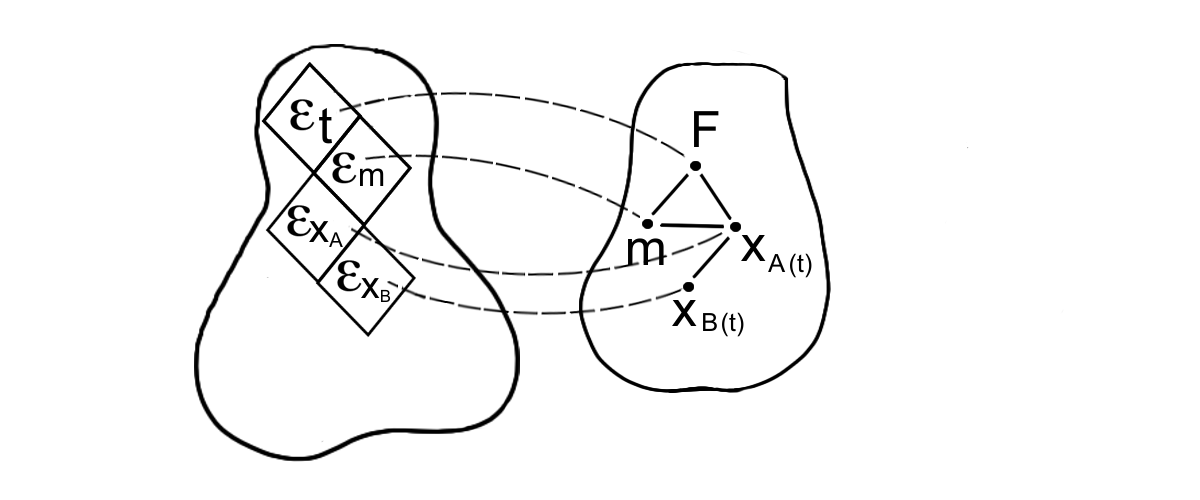}

\caption{Newton's theory mapped to elements of reality}

\end{figure} 

But there is a problem. Coordinates can be transformed to a different reference frame. Let us apply Galilean transformations on the elements of our theory.

\begin{equation}
x^\prime=x(t)+v\cdot t+a
\end{equation}
 
Without going into details, it turns out that force and mass will be invariant under Galilean transformation, but we will receive transformed elements of the theory for $x_A(t)$ and $x_B(t)$. A mapping of the transformed theory $T^\prime$ delivers new elements of reality. 

\begin{figure}[h]
\centering

\includegraphics[width=0.6\textwidth]{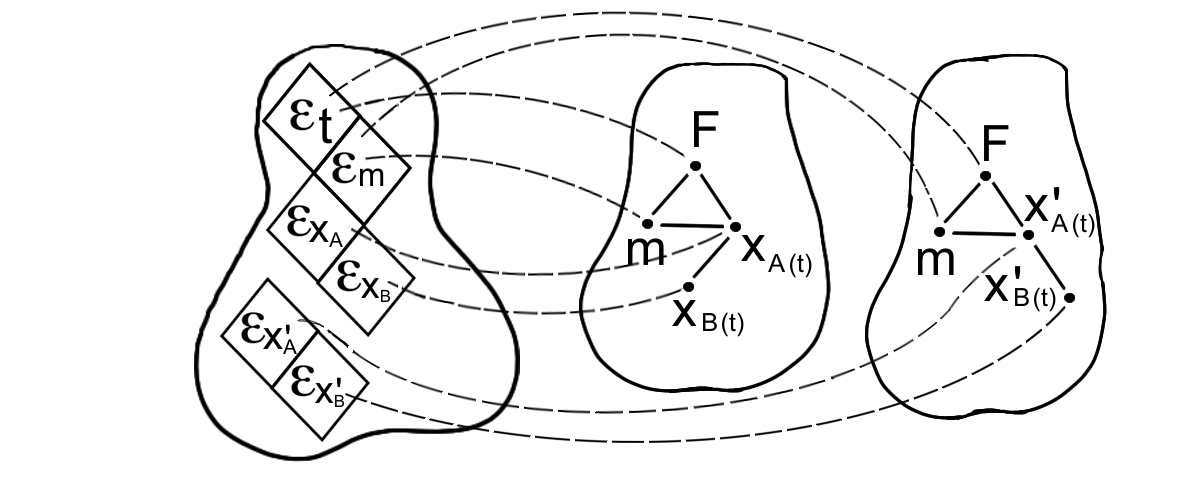}

\caption{Newton's theory mapped to elements of reality}

\end{figure}

This mapping shows us that only by using deterministic elements of the theory, and therefore fulfilling Einstein's criterion of reality, we are not able to identify elements of reality that are really satisfying, because they cannot be identified as unambiguous. 

What can be done to resolve this unsatisfying situation is to choose the elements of the theory differently. The position of the particle A can also be described by the distance to the particle B.

\begin{equation}
d(t)=x_A(t)-x_B(t)
\end{equation}

It turns out that this distance $d(t)$ is invariant under Galilean transformation and the mapping $\phi$ and $\phi^\prime$ of the theory $T$ and its Galilean-transformed alternative $T^\prime$. We therefore obtain elements of reality that are satisfying due to their invariance. 

\begin{figure}[h]
\centering

\includegraphics[width=0.6\textwidth]{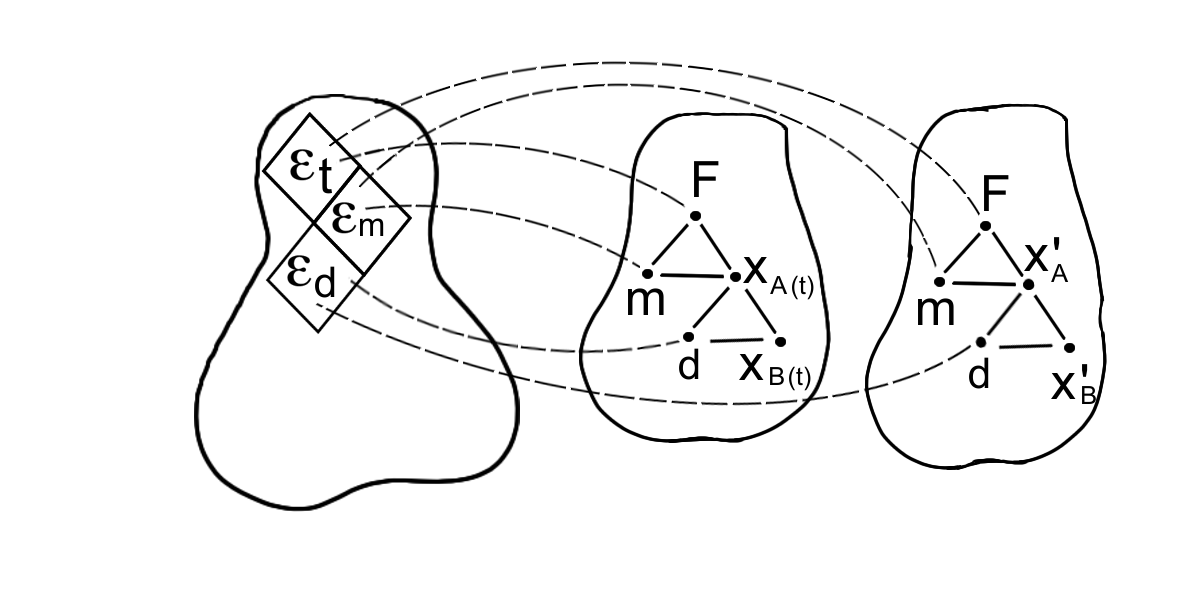}

\caption{Newtons theory mapped to elements of reality with invariants}

\end{figure}

The chosen example was simple to demonstrate this idea, but can be transferred to the theory of relativity or any other classical theory. Especially in the theory of relativity, the concept of invariance(covariance) will play a foundational role in the conception of the theory. Einstein refers to this invariance principle when he states \cite[]{einstein1936physik}:

\begin{quote}
\textit{``Nur solche Gleichungen sind als Ausdruck allgemeiner Naturgesetze sinnvoll,
welche bez\"uglich in diesem Sinne beliebiger Transformationen kovariant sind. (Postulat der allgemeinen Kovarianz).''}
\end{quote}
 
\begin{quote}
\textit{``Only such equations that fulfil covariance regarding any transformation are resaonable to represent laws of nature. (Postulate of general covariance).\footnote{Translation by the author}}''
\end{quote}

The reality criterion is a method to find those invariant elements in the theory to identify corresponding elements of reality that have ontological character. 

Lehner emphasizes the role of invariants in \cite[p.313]{lehner2014einstein}: 

\begin{quote}
\textit{``The association of descriptive invariants with objective reality has become a methodological standard in modern physics: it is the conceptual basis not only of general relativity, but of all forms of gauge theories, also in their quantum forms.''}
\end{quote} 

This key role of invariances will encourage Lehner to define a methodological realism he ascribes to Einstein's philosophy of physics, which we will discuss in section \ref{Section Cornerstones}. \newline

Another aspect of the reality criterion is worth a discussion.
\cite[]{fine2009shaky} states that the reality criterion can only be found in the EPR paper and was not used by Einstein in any other presentation of the EPR argument: 

\begin{quote}
\textit{``...one should note that although Einstein himself later published several versions of EPR, none of them makes any reference to or use of that reality criterion.''}
\end{quote}

We can show that this is not correct. In his letter to Karl Popper, Einstein presented the EPR argument in a compact way, but clearly made use of the reality criterion when he states \cite[]{popper2005logic}:

\begin{quote}
\textit{``For if, upon freely choosing to do so [that is, without interfering with it], I am able to predict something, then this something must exist in reality.''}\footnote{
The original wording is slightly different and we would prefer to give a different translation:
\begin{quote}
\textit{``Denn was ich nach freier Wahl prophezeien kann, das muss auch in der Wirklichkeit existieren''}
\end{quote}

\begin{quote}
\textit{``What can be predicted and freely choosen with conviction, that has to exist in reality.''} 
\end{quote}
}
\end{quote}

The reality criterion is basic to one main agenda in the EPR paper; to criticise the statistical character of Quantum Mechanics.

\newpage

\subsubsection*{The statistical character of Quantum Mechanics}
\label{section statistical character}

In the EPR paper, no direct reference can be found to answer the question if in the EPR argument the quantum-mechanical description refers to a statistical ensemble or to single events, to a single quantum-mechanical object. It is not discussed explicitly, though one can conclude for good reason that a single object is meant when they repeatedly speak about a particle's behaviour \cite[p.779]{einstein1935can}:

\begin{quote}
\textit{``The fundamental concept of the theory is the concept of state, which is supposed to be completely characterized by the wave function $\psi$, which is a function of the variables chosen to describe the particle's behaviour.''}
\end{quote}

\begin{quote}
\textit{``For this purpose let us suppose that we have two systems, I and II,... We can then calculate with the help of Schr\"odinger's equation the state of the combined system I + II at any subsequent time;''}
\end{quote}

Using the term "system," the authors left open whether they apply the quantum-mechanical description to an statistical ensemble or to a single quantum-mechanical object. One year later, after several discussions and after a reply paper by Bohr, Einstein had chosen a clear position on this question \cite[]{einstein1936physik}:

\begin{quote}
\textit{``Es scheint mir deshalb klar, dass die Born 'sche statistische Deutung der Aussagen der Quantentheorie die einzig m\"ogliche ist: Die $\psi$-Funktion beschreibt \"uberhaupt nicht einen Zustand, der einem einzelnen System zukommen k\"onnte; sie bezieht sich vielmehr auf so viele Systeme, eine "System-Gesamtheit" im Sinne der statistischen Mechanik.''}
\end{quote}

\begin{quote}
\textit{``It is obvious to me that the statistical interpretation of quantum theory by Born is the only way: The $\psi$-function does not describe a state that represents one single system; It refers to many systems, to a system-totality in the sense of statistical mechanics.''\footnote{Translation by the author}}
\end{quote}

Even more, he refers to the EPR paper and states that the ensemble interpretation solves any difficulty \cite[]{einstein1936physik}:

\begin{quote}
\textit{``Eine solche Interpretation beseitigt auch eine yon mir zusammen mit zwei Mitarbeitern j\"ungst dargestellte Paradoxie ... Die Zuordnung der $\psi$-Funktion zu
einer System-Gesamtheit beseitigt auch hier jede Schwierigkeit.''}
\end{quote}

\begin{quote}
\textit{``Such an interpretation also removes a recently by me and two colleagues presented paradox ... The assignment of the $\psi$-function to a system-totality here also eliminates all difficulties.''\footnote{Translation by the author}}
\end{quote}

It is not clearly defined what Einstein had in mind with these "difficulties", but it is most likely that he refers to critiques on the EPR paper \cite[]{bohr1935can}; \cite[]{ruark1935quantum}. Without any doubt the difficulties do not refer to the question if Quantum Mechanics is complete or incomplete. That position of Einstein is unchanged as one can see from the presented statement concerning the EPR argument one year later \cite[]{einstein1936physik}:

\begin{quote}
\textit{``Diese Zuordnung mehrerer $\psi$-Funktionen zu demselben physikalischen Zustande des Systems B zeigt wieder, dass die $\psi$-Funktion nicht als (vollst\"andige) Beschreibung eines physikalischen Zustandes (eines Eingelsystems) gedeutet werden kann.''}
\end{quote}

\begin{quote}
\textit{``This assignment of several $\psi$-functions to one physical state of the system B demonstrates that the $\psi$-function cannot be interpreted as a (complete)description of a physical state (of a single quantum mechanical object)''\footnote{Translation by the author}}
\end{quote}

Einstein changed his position on the statistical character on Quantum Mechanics due to critique. Of course, he was aware of the statistical character of Quantum Mechanics before,\footnote{In his letter to Schr\"odinger where he explained the EPR argument, he also refers to the statistical character\cite[letter 206,p.537]{vonMeyenn2010entdeckung}:  
\begin{quote}
\textit{``Nun beschreibe ich einen Zustand so: Die Wahrscheinlichkeit dafür, daß die Kugel in der ersten Schachtel ist, ist $\frac{1}{2}$
Ist dies eine vollst\"andige Beschreibung? 
Nein: Eine vollst\"andige Aussage ist: die Kugel ist in der ersten Schachtel (oder ist nicht). So muß also die Charakterisierung des Zustandes bei vollst\"andiger Beschreibung aussehen.''}
\end{quote}
\begin{quote}
\textit{``I describe a state this way: The probability for a ball being in the first box is $\frac{1}{2}$. Is that a complete description? 
No: A complete description is: The ball is in the first box (or is not). The representation of a complete state has to be that way.'' Translation by the author.}
\end{quote}
} but the Born interpretation was an argument for incompleteness that was much easier to defend than the EPR argument. \newline

%
%

The statistical character of Quantum Mechanics is connected to the notion of completeness. We will see in \ref{SectionCompleteness} that there are several definitions of completeness and there is a deep connection to the statistical character of Quantum Mechanics. Before discussing the connection between EPR and ER paper, and based to the previously presented aspects, before we will analyse cornerstones of Einstein's realism.

\newpage

\subsubsection*{Cornerstones of Einstein's realism}
\label{Section Cornerstones}
The term reality cannot be seen isolated; it is connected to several conceptions that accompany reality. Neither can a conception of reality of a scientist be seen as unvarying; it is subject to change and variation and often inconsistent\footnote{By this inconsistency we mean the contradiction between several ontological statements or concepts that refer to reality.}. A thorough discussion of Einstein realism would go beyond the scope of this paper; therefore, we will only present aspects or cornerstones of the ideas of reality that Einstein had in mind. For a more detailed analysis of the evolution of Einstein's philosophical position over the years, refer to \cite[]{howard1993einstein}, an in-depth discussion on Einstein's view on realism and philosophy of science is given by \cite[]{janssen2014cambridge}.


\begin{itemize}

\item Objectivity - There is a world outside, independent of our experiences, descriptions and ideas, a given reality that awaits to be uncovered. Connected to that, Einstein also has in mind that there is a description that reflects this truth about nature. In that sense, this view is contradictory in its claim to represent elements of the theory as human thoughts and ideals and simultaneously follow a correct path, as Einstein calls it. See section \ref{section view on theories} "Einstein's view on theories and theory formation".

\item (Mathematical) - Simplicity: Einstein has in mind that a physical theory has to fulfill simplicity \cite[p.165]{einstein1934method}:

\begin{quote}
\textit{``It can scarcely be denied that the supreme goal of all theory is to make the irreducible basic elements\footnote{By elements he refers to the Machian elements of the theory.} as simple and as few as possible without having to surrender the adequate representation of a single datum of experience.''}
\end{quote}

Moreover, he assumed this simplicity is something that is realised in nature as well \cite[p.165]{einstein1934method}:

\begin{quote}
\textit{``Our experience up to date justifies us in feeling sure that in Nature
is actualized the ideal of mathematical simplicity.''}
\end{quote}

This is a contradiction in the understanding of simplicity, because the ideas and elements of the theory should fulfill simplicity, but they are merely a construction by human minds, so this simplicity is nothing fundamental and cannot be the same simplicity he mentions in the second quotation. We presented a discussion of the concept of simplicity in context with interpretations of Quantum Mechanics and its peculiarities in \cite[]{krizek2017ockham}. For an account on Einstein's conception of simplicity, refer to \cite[p.230]{howard1993einstein}.

\newpage
 
\item Mathematical entities - Einstein favours the four dimensional continuum over algebraic structures with arbitrary or even infinite dimensionality. Out of some statements, one could conclude that he took the position of naive realism in a sense of taking mathematical entities as elements of reality, a mathematical realism. 

\item Methodological and phenomenal realism - According to \cite[]{lehner2014einstein}, Einstein advanced a view on the role of invariants in theoretical models; he calls that position methodological realism. The role of invariants in a theory is a distinguished one; the invariants represent elements of the theory that are assumed to have a strong connection to objective reality. This conception is based on the theory of relativity, but also found application in modern theoretical physics like in Gauge theories. In contrast to methodological realism, he sees phenomenal realism \cite[p.313]{lehner2014einstein}: 

\begin{quote}
\textit{``By contrast, physical measurements do not reveal those objective entities themselves, but their representations in local coordinate frames, that is, noninvariant quantities. Such measurements are no less real, but they are real in a different sense: their reality is \textit{phenomenal}, dependent on an observer situated in a specific way.''}
\end{quote}

Following Lehners reading, it was Einstein's attempt to apply methodological realism to Quantum Mechanics, and the failure of this attempt was reason for rejecting Quantum Mechanics as a foundational theory.  

\item Deterministic character - Einstein was very clear about that point. To him, a theory with statistical character is only provisional and cannot represent real events. Refer also to the discussion in section \ref{section statistical character}.

\item Non-localizability: In some sense, this non-localizability is a nonlocality feature, but not in the sense of an action-at-a-distance; more in the sense of the notion of a non-absolute localization of particles. The concept of non-localization was familiar to Einstein from the way the gravitational energy behaves in general theory of relativity. We will give a more detailed discussion of Non-localizability in section \ref{localizability}. 

\item Singularity-free - To Einstein the representation of particles in a field theory implied the problem of singularities of the field. To him a reasonable representation of particles has to be free from such singularities in the field. Refer also to the discussion in section \ref{localizability}.

\item Separability: The physical reality allows separation of the world into parts that can be handled independently. To Einstein, this is a prerequisite for scientific reasoning per se (See Don Howard \cite[p.234ff]{howard1993einstein} and \cite[p.39]{howard2005albert}). The spatiotemporal separability consists of locality and separability of systems, per se. There is also a separability without a spatiotemporal aspect, which is realised, for instance, by linear systems that can be decomposed into systems that evolve independently of each other, even if they share a position in spacetime. Fourier analysis is a good example of this superposition principle that allows decomposition of systems into separable subsystems. 


\item Locality and causality - In Einstein's view, reality obeys locality in the sense of local realism: Events in two spacelike separated areas cannot affect each other. Therefore, all meaningful descriptions of nature have to obey local realism as well. 

\item Unambiguity or uniqueness: A complete description\footnote{See section \ref{SectionCompleteness} for an overview on several definitions of completeness. This completeness will be defined as \textit{theory completeness} and will refer to the elements of the theory.} leaves no room for ambiguity. This is connected to Aristotelian ideas of physics - basically the contradiction principle - which is based on common sense experience. This view reflects also in the EPR argument, in the condition of completeness \cite[]{einstein1935can} and in later variations of the EPR argument \cite[p.537]{vonMeyenn2010entdeckung}; \cite[p.482]{popper2005logic}; \cite[p.341]{einstein1936physik}; \cite[]{einstein1948quantum}.

\end{itemize}

\newpage
\section{EPR=ER in 1935 and its revival}

In this chapter, we provide an answer to the question of whether there is a connection between the EPR and ER paper and, if yes, in what kind of way. In a modern context, where in some interpretations of Quantum Mechanics nonlocality plays a crucial role, and the imprint by popular culture, it is a tempting thought to put the nonlocality features of Quantum Mechanics in connection to nonlocal structures like wormholes. It is even more surprising to find out that the first wormhole paper\footnote{Before the ER paper, only speculations by Flamm and Weyl can be found. As pointed out by \cite[]{visser1996lorentzian} which coined the term "first wormhole paper".} was written by the same persons that challenged the completeness of Quantum Mechanics with the EPR argument, namely Einstein and Rosen. The conclusion one could draw at first glance is close-at-hand: Einstein and Rosen could have had in mind to solving the issue with nonlocalities in Quantum Mechanics by wormholes. This conclusion is wrong due to several reasons we present now. 

\subsection{Nonlocality}

At first, one has to say that there was no issue with nonlocality to solve; at least, not for Einstein. Einstein recognised the nonlocality of Quantum Mechanics, which basically occurred in context with the projection postulate, but to him it was untenable to accept nonlocalities in physics, so he was not searching for a way to explain nonlocalities by an underlying theory. It is this following statement that makes clear that nonlocality, for the authors of the EPR-paper, and it can be assumed especially for Einstein, was not considered to be a possible or even thinkable part of ontology \cite[]{einstein1935can}: 

\begin{quote}
\textit{``This makes the reality of P and Q depend upon the process of measurement carried out on the first system, which does, not disturb the second system in any way. No reasonable definition of reality could be expected to permit this.''}
\end{quote}

The action-at-a-distance effects in Newtonian mechanics were already a problem to Newton and entail a lot of philosophical problems, such as difficulties with causality. It is comprehensible that Einstein was not considering to give up the beautiful new and self-consistent formulation of a causal structure. To him, there must have been a different solution that would integrate the quantum effects in the framework of field theory with its complete\footnote{In the sense of \textit{theory completeness}} character. But field theory had its problems as well, and Einstein was full aware of them. 

If nonlocality was not the pursued goal of the ER-paper, what was it then? In his paper on the method of theoretical physics, he clarifies one of the main difficulties in field theory \cite[]{einstein1934method}:

\begin{quote}
\textit{``The most difficult point for such a field-theory at present is how to include the atomic structure of matter and energy. For the theory in its basic principles is not an atomic one in so far as it operates exclusively with continuous functions of space, in contrast to classical mechanics whose most important feature,
the material point, squares with the atomistic structure of matter.''}
\end{quote}  

The absolute localization of particles was an open question to Einstein, and we will see the connection to the ER-paper in section \ref{localizability}.

\subsection{Localizability of particles}
\label{localizability}

In section \ref{section view on theories} we presented what Einstein understood as the term ``atomic structure'': The singularity-free representation of particles in space. To him, it was not only a problem of field theory, but had a deep connection to quantum theory \cite[]{einstein1934method}: 

\begin{quote}
\textit{``... it seems to me certain that we have to give up the notion of an absolute localization of the particles in a theoretical model. This seems to me to be the correct theoretical interpretation of Heisenberg's indeterminacy relation.''}
\end{quote} 

Not only that the particle is represented singularity-free, the refutation of an absolute localization of the particle was the goal of this concept\footnote{It is noticeable that this highly non-classical idea has been proposed by Einstein, who repeatedly had to face critique on his inflexible position and his adherence to classical physics.}. To Einstein, it promised the reconstruction of quantum effects by general theory of relativity. 

The ER-paper opens in the abstract with reference to this atomistic theory of matter \cite[]{einstein1935particle}: 

\begin{quote}
\textit{``The writers investigate the possibility of an atomistic theory of matter and electricity which, while excluding singularities of the field, makes use of no other variables than the $g_{\mu\nu}$ of the general relativity theory and the $\phi_{\mu\nu}$ of the Maxwell theory.''}
\end{quote} 


In the model of the Einstein-Rosen bridge, the bridge is a representation of a particle, but the position of the particle is smeared out. The position is in contrast to the external Schwarzschild-solution not given by the position coordinate of the singularity. The metric around the particle is defined by the Einstein-Rosen-bridge solution; still, there is no account of the particles exact position. This is satisfying in the sense of Einstein's requirement for solving the quantum problem within a continuum theory presented in section \ref{section view on theories} \cite[]{einstein1934method}. \newline

It is comprehensible that the goal of the ER-paper was to present a field theory that solves the localisation problem for particles. Even more, is was pursued to reconstruct quantum theory with a unified field theory that can claim to be complete. The definition of completeness and its peculiarities will be presented in section \ref{SectionCompleteness}.  

For a more detailed discussion concerning the philosophical implications of the concept of particles, refer to \cite[]{achinstein1991particles} and \cite[]{kuhlmann2010ultimate}.

\newpage
\subsection{Completeness}

\label{SectionCompleteness}

The completeness used in the EPR argument, ER paper and accompanying sources is a peculiar concept. At first hand, one has a clear idea what is understood by completeness, but we will see that a clear definition is needed for the discussion of the context on the EPR argument. We will show that there are several definitions of completeness involved. The connection to the ER-paper will be presented. 

\subsubsection{Theory completeness}

In 1934, Einstein defined the meaning of the term completeness \cite[]{einstein1934method}: 

\begin{quote}
\textit{``A complete system of theoretical physics consists of concepts and basic laws to interrelate those concepts and of consequences to be derived by logical deduction.''}
\end{quote}

In section \ref{section Classification scheme} a notion of the structure of a physical theory has been presented. By using physical quantities, laws to connect them, and concepts to give structure and principles to the theory, a complete theory in this sense can be pictured as following:

\begin{figure}[h]
\centering

\includegraphics[width=0.4\textwidth]{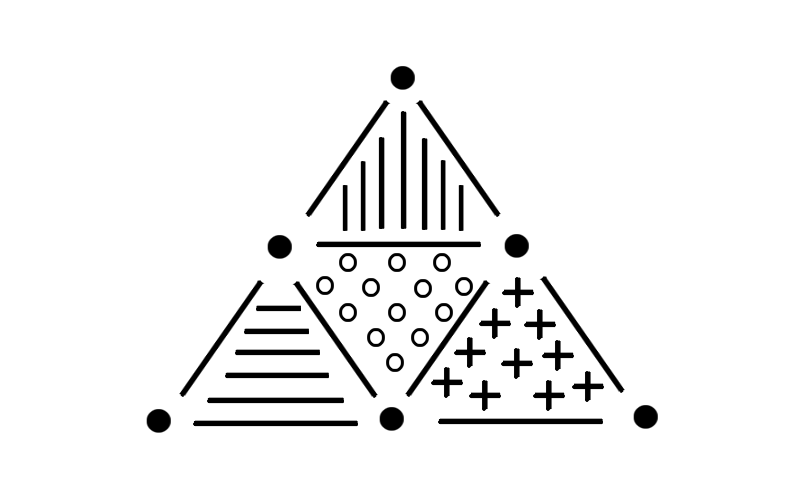}

\caption{Structure of a theory}
\end{figure}

The mentioned elements (physical quantities, laws and concepts/principles) are meant in the Machian sense and not connected, per se, to an ontology. Einstein emphasized that these elements are ``...free inventions of the human mind...'' \cite[]{einstein1934method} and therefore pure constructions of thought. Therefore, this definition of completeness can be understood more as a consistency in the theoretical system. In any sense, it forgoes to provide ontological statements. We will call this completeness \textit{theory completeness}. \newline 

An application of this theory completeness is given in \cite[]{einstein1935particle}:

\begin{quote}
\textit{``One of the imperfections of the original relativistic theory of gravitation was that as a field theory it was not complete; it introduced the independent postulate that the law of motion of a particle is given by the equation of the geodesic.
A complete field theory knows only fields and not the concepts of particle and motion. For these must not exist independently of the field but are to be treated as part of it.''}
\end{quote}

We want to give a definition for theory completeness that gets as close as possible to the concept Einstein had in mind:

\begin{mydef}
A theory fulfills theory completeness if its physical quantities and laws are connected by concepts that relate only to each other, and the theory renounces extra assumptions of concepts that exceed the framework of the theory. 
\end{mydef}

\textit{Theory completeness} does not refer to an ontology. It only refers to elements of the theory, and requests a consistency in the used concepts. Einstein refers to it in \cite[]{einstein1935particle} as Closeness(Vollst\"andigkeit).

\begin{figure}[h]
\centering

\includegraphics[width=0.6\textwidth]{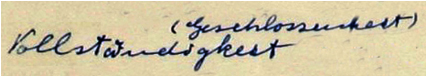}

\caption{Completeness - Closeness \cite[]{einstein1935DasPartikelProblemHandschrift}}
\end{figure}


References to \textit{theory completeness} can be found in: \cite[p.164,166]{einstein1934method}; \cite[p.778]{einstein1935can}; \cite[p.76, p.77]{einstein1935particle}; \cite[p.316, p.344]{einstein1936physik}; \cite[p.320, p.323]{einstein1948quantum}.

\subsubsection{Bijective completeness}

The EPR paper posed a question on completeness, but is it \textit{theory completeness} that is challenged? If we refer to the EPR paper, a definition of completeness is given \cite[]{einstein1935can}:

\begin{quote}
\textit{``Whatever the meaning assigned to the term complete, the following requirement for a complete theory seems to be a necessary one: every element of the physical reality must have a counterpart in the physical theory''}
\end{quote}

\begin{figure}[h]
\centering

\includegraphics[width=0.6\textwidth]{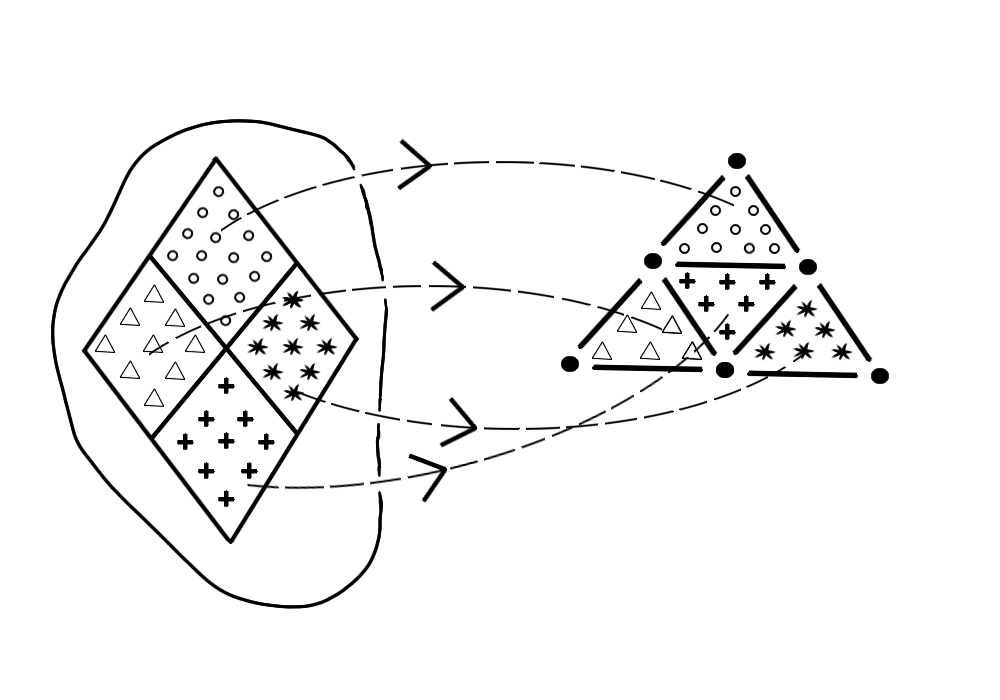}

\caption{Bijective completeness}
\end{figure}

As can be seen immediately, this definition of completeness differs from the definition of \textit{theory completeness}; it even further extends the term completeness to the ontological level, whereas the \textit{theory completeness} definition refers only to the theory and its elements. The latter relates the elements of reality to the elements of a theory. \cite[]{fine2009shaky} calls this completeness \textit{bijective completeness}. Following \cite[]{einstein1935can}, a definition would be: 

\begin{mydef}
\label{Definition bijective completeness}
Bijective completeness means that every element of the physical reality $\epsilon_i $ must have a counterpart  $ t_i $ in the physical theory.
\end{mydef}

Though \textit{bijective completeness} is a Bijection between elements of the theory and elements of reality, it starts with the latter ones. Starting at elements of reality causes the problem how of to identify them and the corresponding elements of the theory. For this purpose, \cite[]{einstein1935can} had to assume a criterion of reality to identify these elements of reality\footnote{See also Section \ref{Section Elements of reality}.}. \newline 

From an accompanying source, one more aspect about \textit{bijective completeness} can be seen. Karl Popper received a letter from Einstein in September of 1935, where Einstein explained the idea of the EPR paper in a short and very clear way. One more aspect of Einstein's notion of \textit{bijective completeness} can be found there \cite[p.413]{popper2005logic}:

\begin{quote}
\textit{``Since a complete description of a physical state must necessarily be an unambiguous description (apart from superficialities such as units, choice of the co-ordinates etc.), it is therefore not possible to regard the $\psi$-function as the complete description of the state of the system.''}
\end{quote}

He concludes that a \textit{bijective completeness} description has to be unambiguous, so if every element of the physical reality has to have a (one) correspondence in the theory, it must be vice versa. This unambiguousness assumption is not presented explicitly as with the reality criterion, the completeness criterion, or separability/locality, though it plays an important role in the EPR argument. 

\begin{figure}[h]
\centering

\includegraphics[width=0.6\textwidth]{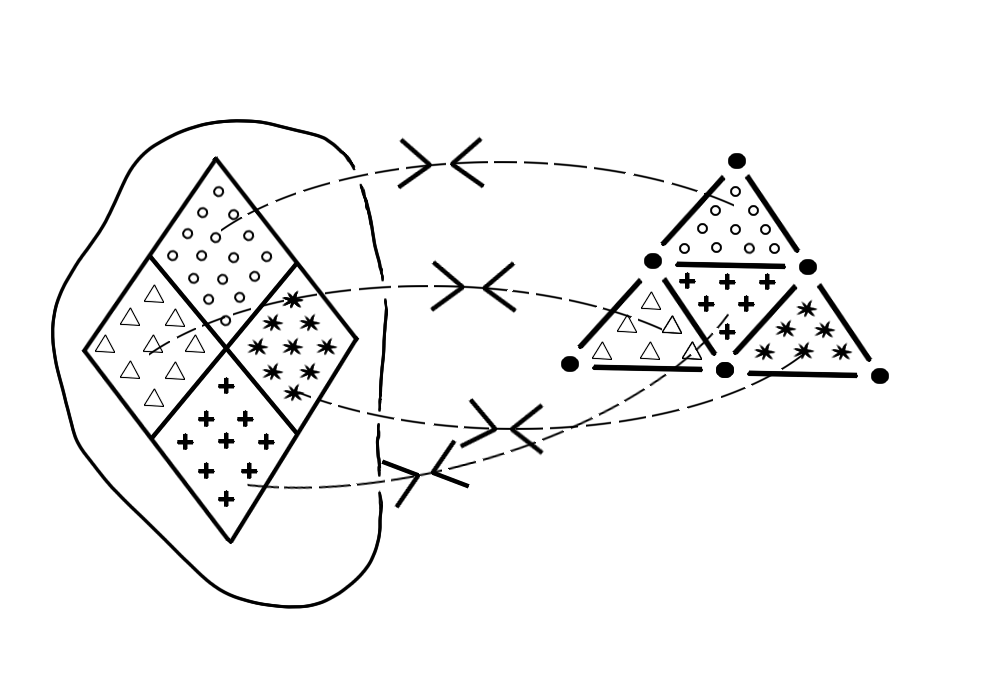}

\caption{Bijective completeness and unambiguousness}
\end{figure}

It is further interesting that Einstein seems to have seen a problem with the pure \textit{bijective completeness} argument in the EPR paper. Superficialities as choice of coordinates, as we have presented them before in section \ref{Section Elements of reality} would not fulfill \textit{bijective completeness}, so Einstein clarifies this to Popper \cite[p.413]{popper2005logic}. \newline

Einstein gives no exact definition of completeness in the EPR paper, but states that definition \ref{Definition bijective completeness} of \textit{bijective completeness} is a necessary requirement for his concept of completeness. Is it possible that \textit{theory completeness} is the completeness Einstein had in mind? \textit{Theory completeness} is only defined by elements of the theory, and makes no statement about the ontology of the theory; therefore it can be ruled out that \textit{bijective completeness} is a necessary requirement of \textit{theory completeness}. 
\textit{Bijective completeness} would have to be true, so that \textit{theory completeness} could be true, but it is obvious that the structure of the elements of a theory is completely independent of any mapping $\phi$ between the elements of the theory and elements of reality. \textit{Theory completeness} is not the completeness Einstein had in mind, for which he used \textit{bijective completeness} as a necessary requirement. \newline

References to the \textit{bijective completeness} can be found in \cite[p.777, p.778, p.780]{einstein1935can} and \cite[p.341]{einstein1936physik}.



\subsubsection{Born completeness}

Since \textit{theory completeness} is excluded from being the completeness meant in the EPR paper, there must be a different definition of completeness. \cite[]{fine2009shaky} identifies two further concepts of completeness: \textit{Born completeness} and \textit{Schr\"odinger completeness}: \newline 

\begin{quote}
\textit{``According to the Bornian conception, a complete description is essentially nonprobabilistic; genuinely probabilistic assertions are necessarily incomplete.''}
\end{quote}

These conceptions of completeness can be identified in letters from Einstein to Schr\"odinger, where Einstein also stated the "Einstein's boxes" Gedankenexperiment, which discusses the probabilities of finding a ball in a box \cite[letter 206, p.537]{vonMeyenn2010entdeckung}.

\begin{mydef}
\label{Definition Born completeness}
Born completeness for a theory is fulfilled if the theory contains no probabilistic elements. A theory which contains statistical accounts and probabilistic statements is considered to be incomplete. 
\end{mydef}

Following this definition, Quantum mechanics is incomplete per definition. No further arguments are needed. \newline 

Is \textit{Born completeness} the completeness Einstein meant in the EPR paper? If \textit{Born completeness} would be addressed, it would mean that the authors of the EPR paper had in mind that without \textit{bijective completeness}, \textit{Born completeness} could not be true. That follows of the EPR condition of completeness.

On the other hand, the statement ``Quantum mechanics fulfills \textit{Born completeness}'' is false. In this case, out of the rules for the material conditional, it follows that independent of the truth value of the \textit{bijective completeness}, the statement ``\textit{theory completeness} is a necessary requirement for \textit{Born completeness}'' is true. To show the truth value of \textit{bijective completeness} is therefore meaningless, and we can rule out that \textit{Born completeness} is meant as the conception of completeness in the EPR paper. \newline

References to the \textit{Born completeness} can be found in \cite[p.339, p.341, p.343, p.347]{einstein1936physik}.

\subsubsection{Schr\"odinger completeness}

\cite[p.71]{fine2009shaky} gives an analysis of \textit{Schr\"odinger completeness}: 

\begin{quote}
\textit{``By contrast the Schr\"odinger view is that probabilities can be fundamental, not to be reduced to something else. Thus the Schr\"odinger conception is that a complete description of a state of affairs can be a probabilistic assertion, with probability less than unity, which (somehow) tells the whole truth about that state of affairs. If there were some further truth to be told, then the probabilistic assertion would be an incomplete description''}
\end{quote}

It's getting a bit vague with \textit{Schr\"odinger completeness} because it seems that Einstein's conception of it is not a very clear one; at least, he did not define it explicitly. What can be seen immediately is that the definitions of probability in context with \textit{Born completeness} and  \textit{Schr\"odinger completeness} differ vastly. Whereas the concept of the probability of occurrence of outcomes in an ensemble of events is used in the first conception, the latter defines probability as fundamental and related to single events. The interpretation of probability will go beyond the scope of this work. We will therefore stick to an operational definition of \textit{Schr\"odinger completeness} that is based on Einstein's considerations in his letter to Schr\"odinger \cite[letter 206, p.537]{vonMeyenn2010entdeckung}:
\newpage

\begin{mydef}
\label{Definition Schrodinger completeness}
Schr\"odinger completeness for a theory is fulfilled if the theory contains probabilistic statements about elements of the theory. The probabilistic statements refer not to the statistical ensemble of many events, but to one single event. The probabilistic statements are fundamental in the sense that nothing more can be said about the system. 
The theory is considered to be incomplete if there are potential new elements\footnote{Einstein calls them "fremde Faktoren"(external factors). The translation of "fremd" is peculiar, since in the literal translation it would be "strange", "weird", "alien" or "different." But in context, it refers to factors that are not considered in the theory. They are not part of the theory, so we decided to use the non-literal translation "external." In our modern context, these factors could be understood as "hidden variables."} in the theory that would abandon the probabilistic character of the theory. 
\end{mydef}

The definition of \textit{Schr\"odinger completeness} refers to what we would nowadays call "hidden variables." 

It is difficult to decide if \textit{Schr\"odinger completeness} is the completeness Einstein is referring to in the EPR paper. If we carefully analyse the argument in the letter to Schr\"odinger from the 19th of June 1935, we find a self-reference in the definition of completeness \cite[letter206, p.537]{vonMeyenn2010entdeckung}:

\begin{quote}
\textit{``Nun beschreibe ich einen Zustand so: Die Wahrscheinlichkeit daf\"ur, daß die Kugel in der ersten Schachtel ist, ist $\frac{1}{2}$
Ist dies eine vollst\"andige Beschreibung? 
Nein: ...
Ja: ... Der Zustand vor dem Aufklappen ist durch die Zahl $\frac{1}{2}$
vollst\"andig charakterisiert, deren Sinn sich bei Vornahme von Beobachtungen allerdings nur als statistischer Befund manifestiert.''}
\end{quote}

\begin{quote}
\textit{``I describe a state this way: The probability for a ball being in the first box is $\frac{1}{2}$. Is that a complete description? 
No: ...
Yes: ... The state before opening the box is completely characterised by the number $\frac{1}{2}$, whose meaning becomes manifest as statistical account by applying measurements on the system.}
\end{quote}

What is meant by this complete characterisation by the number $\frac{1}{2}$? If we assume it to be \textit{bijective completeness} in a way that the number is ascribed to an element of reality, and the number is interpreted as probability in an epistemic way, this would represent an epistemic account of the situation. Depending on the interpretation of probability, it would represent propensity, knowledge about the state of the ball, or degrees of belief on its state.  \newline

In principle, this would be a description that would fulfill \textit{bijective completeness}, because one can give a Bijection to elements of reality in each of these accounts. Therefore, it is plausible to identify \textit{Schr\"odinger completeness} as the completeness Einstein challenged in the EPR paper. 

\newpage

\subsubsection{$\psi$ - completeness}

In an account to classify hidden variable models, \cite[]{spekkensharrigan2010einstein} proposed a definition of \textit{$\psi$-completeness}:

\begin{quote}
\textit{``An ontological model is $\psi$-complete if the ontic state space $\Lambda$ is isomorphic to the projective Hilbert space $\mathcal{PH}$ (the space of rays of Hilbert space) and if every preparation  procedure $P_\psi $ associated  in  quantum theory with a given ray $\psi$ is associated in the ontological model with a Dirac delta function centered at the ontic state $\Lambda$ ψ that is isomorphic to $\Psi$, $p(\lambda \vert P_\psi)=\delta(\lambda-\lambda_\psi)$.''}
\end{quote}

This definition of \textit{$\psi$-completeness} is claimed to be identical with \textit{bijective completeness} \cite[p.147]{spekkensharrigan2010einstein}:

\begin{quote}
\textit{``It is quite clear that by "real state of the real system", Einstein is referring to the ontic state pertaining to a system.  Bearing this in mind, his definition of completeness can be identified as precisely our notion of $\psi$-completeness given in Definition 2''}
\end{quote}

This is a misunderstanding of Einstein's view on completeness and elements of reality. Einstein is not using the wording "elements of reality" in his letter to Schr\"odinger\cite[letter206, p.538]{vonMeyenn2010entdeckung}, but referring to \textit{bijective completeness}:

\begin{quote}
\textit{``Man beschreibt in der Quantentheorie einen wirklichen Zustand eines Systems
durch eine normierte Funktion $\psi$ der Koordinaten (des Konfigurationsraumes).Die
zeitliche \"Anderung ist durch die Schr\"odinger-Gleichung eindeutig gegeben. Man
m\"ochte nun gerne folgendes sagen: $\psi$ ist dem wirklichen Zustand des wirklichen
Systems eindeutig zugeordnet. Der statistische Charakter der Me\ss ergebnisse
f\"allt ausschließlich auf das Konto der Messapparate bzw. des Prozesses der Messung.
Wenn dies geht rede ich von einer vollst\"andigen Beschreibung der Wirklichkeit
durch die Theorie. Wenn aber eine solche Interpretation nicht durchf\"uhrbar ist,
nenne ich die theoretische Beschreibung „unvollst\"andig“.''}
\end{quote}

\begin{quote}
\textit{``In quantum theory a real state of a system is described by a normed function $\psi$ of coordinates (of configuration space). The evolution in time is given by Schr\"odinger equation uniquely. One would like to say: $\psi$ is corresponding to the real state of the real system uniquely. The statistical character of the measurement results is a consequence of the measurement apparatus respectively the measurement process. If this works, I speak about a complete description of reality by a theory. If such an interpretation is not possible, I call the theoretical description incomplete.\footnote{Translation by the author}''}
\end{quote}

\cite[]{spekkensharrigan2010einstein} conclude out of these statements that Einstein's \textit{bijective completeness} is identical to their \textit{$psi$-completeness}, but
the latter one is setting up an Isomorphism between the projective Hilbert space and an ontic state space, whereas \textit{bijective completeness} is a Bijection between elements of a theory and elements of reality. These elements of reality have no defined mathematical structure, per se. In our proposed classification scheme for interpretations of Quantum Mechanics, the elements of reality are located in level four, the ontological level. Assuming that the elements of reality are elements of an ontic state space, is a kind of mathematical realism, but we think that is not the intention of the claim. \newline

On the other hand, \cite[]{spekkensharrigan2010einstein} have good reasons to assume that Einstein, by referring to the "real state of the real system" \cite[letter 206, p. 538]{vonMeyenn2010entdeckung}, means elements of a theory that would replace Quantum Mechanics. They propose an argument for Einstein's preference for the EPR argument, in contrast to his incompleteness argument from 1927. One aim, according to them, was Einstein's goal to emphasize the epistemic character of the state vector and to look out for hidden variable theories.  \newline 

\textit{$\psi$-completeness} is a useful classification tool for hidden variable theories, but we refuse the identity of \textit{$\psi$-completeness} and \textit{bijective completeness} based on the involvement of completely different structures in these two conceptions. This does not affect any applications of \textit{$\psi$-completeness} in \cite[]{spekkensharrigan2010einstein}.

\subsubsection{Standard completeness}

The following completeness criterion is not part of the original EPR discussion, and is presented here to amend the overview on accounts on completeness.  
\cite[]{held2012incompatibility} gives an argument against \textit{standard completeness}. He brings the timeframe of the involved events into consideration and claim a conflict between \textit{standard completeness} and the principle that "QM should deliver probabilities for physical systems possessing properties at definite times" \cite[]{held2015einstein}. 

\subsubsection{Remarks on completeness and the EPR-ER connection}

The targets of the EPR paper are manifold; one aim is to show that  \textit{Schr\"odinger completeness} does not hold for the unitary core of Quantum Mechanics itself. Einstein claimed that Quantum Mechanics did not receive its statistical character out of the measurement process. The unitary evolution of the state vector itself is, according to him, incomplete in the sense of \textit{Schr\"odinger completeness}. The critique on Quantum Mechanics is one story, but more interesting in our context is the motivation for showing this incompleteness. Replacing a successful theory with a new unknown theory without a good reason encounters opposition. Einstein had to convince the community of the incompleteness of Quantum Mechanics to present a new program \cite[]{einstein1935can}: 

\begin{quote}
\textit{``While we have thus shown that the wave function does not provide a complete description of the physical reality, we left open the question of whether or not such a description exists. We believe, however, that such a theory is possible.''}
\end{quote}

Einstein retreated quickly, already in 1936, to the position of \textit{Born completeness}, which to him was sufficient to show the need for a theory that fulfills his requirements. Referring to his article \cite[]{einstein1948quantum}, he notes in a letter to Besso \cite[p.403]{speziali1972correspondance}: 

\begin{quote}
\textit{``Es freut mich dass Du meinen kleinen Aufsatz gelesen hast. Hast Du auch gemerkt wie unlogisch Pauli darauf geantwortet hat? Er leugnet es, dass diese Art der Beschreibung unvollst\"andig sei, sagt aber im selben Atemzuge, dass die $\psi$ Funktion eine statistische Beschreibung des Systems sei, die Beschreibung einer System-Gesamtheit. Dies ist doch nur eine andere Form der Aussage. Die Beschreibung des (individuellen) Einzelsystems ist unvollst\"andig!  ''}
\end{quote}

\begin{quote}
\textit{`` I am pleased to hear that you read my little article. Have you recognised also how illogical Pauli responded? He denies that this description is incomplete, but states in the same breath that the $\psi$ function is a statistical description of the system, the description of a system-totality. This is just a different formulation of the statement: The description of an (individual) single-system is incomplete!\footnote{Translation by the author}''}
\end{quote}

This statement is the essence of \textit{Born completeness}. And it shows that the EPR arguments and all its successor arguments by Einstein had one main purpose: To show the incompleteness of Quantum Mechanics for the specific purpose of justifying the need for a new field theory to replace Quantum Mechanics.

Despite the ambiguous notions of completeness, it is fair to conclude that it was the idea to present this new complete theory in the next published work following two months after the EPR paper; the Einstein-Rosen bridge paper. \newline

We presented the meaning of the terms "completeness" and "reality" and pointed out which aspects of reality were specially relevant to Einstein in context with the 1935 papers. In the following chapter, we will make use of these clarifications to discuss the connection of the 1935 papers in the context of the unified field theory program. 

\newpage

\subsection{The Unified field theory program}

Einstein had in mind to resolve the incomplete\footnote{The completeness refers to \textit{Born completeness}} statistical character of quantum theory with a unified field theory. Even in 1934, one of the conclusive statements of his paper on the method of theoretical physics addresses the statistical character \cite[]{einstein1934method}: 

\begin{quote}
\textit{``I still believe in the possibility of giving a model of reality, a theory, that is to say, which shall represent events themselves and not merely the probability of their occurrence.''}
\end{quote}

We have seen before that to Einstein the concept of completeness is strongly connected to the statistical character of quantum theory. In the original EPR argument in \cite[]{einstein1935can}, he used the reality criterion, locality(Trennungsprinzip - separation principle), and the contradiction principle to conclude the incompleteness of Quantum Mechanics. In \cite[]{einstein1936physik}, he no longer referred to those principles, instead, he argued for the ensemble interpretation, which is an indirect use of the \textit{Born completeness}. According to the ensemble interpretation, Quantum Mechanics only refers to a statistical ensemble to events, not to a single event itself. To Einstein, this argument was sufficient to show the incomplete character of Quantum Mechanics and to head for an underlying theory. We can see that this is one of the main agendas of the EPR paper \cite[]{einstein1935can}:   

\begin{quote}
\textit{``While we have thus shown that the wave function does not provide a complete description of the physical reality, we left open the question of whether or not such a description exists. We believe, however, that such a theory is possible.''}
\end{quote}

This new theory he presented in the ER-paper, at first independent of the EPR critique, was meant to solve the problem of the atomistic character of matter and fulfils Einstein's requirement of completeness \cite[]{einstein1935particle}: 

\begin{quote}
\textit{``In favor of the theory\footnote{The Einstein-Rosen bridge model for elementary particles} one can say that it explains the \underline{atomistic} \underline{character} of matter as well as the circumstance that there exist no negative neutral masses, that it introduces no new variables other than the $g_{\mu\nu}$ and $\phi_{\mu\nu}$, and that in principle it can claim to be \underline{complete} (or closed). On the other hand one does not see a priori whether the theory contains the quantum phenomena.'' }[our underlining]
\end{quote}

Here he means \textit{theory completeness}, which is evident when he puts the term "closed" in brackets to underline that it refers to the closeness of a unified field theory that hopefully solves the particle problem. \newline

It is remarkable that Einstein differs between the atomistic character of the theory and the question if the theory contains quantum phenomena. We have seen before what he means exactly by ``atomistic character'', it refers to the problem of the localization of particles without accepting singularities of the field. The Einstein-Rosen bridge would have solved this in a way satisfactory to Einstein. It is not exactly clear what Einstein understands by the term ``quantum phenomena'' and how the theory could contain them. It is comprehensible that he means the quantum states and the formalism of Quantum Mechanics, which cannot be found in the Einstein-Rosen bridge representation for a particle, but it certainly was Einstein's hope that the quantum effects would emerge out of this new model. 

In \cite[]{einstein1936physik}, he presented the EPR argument in a modified form in line with the ER-bridge. Here, the connection of these two papers as part of a program is evident: 

\begin{quote}
\textit{``Es wird aber gezeigt, dass die \"Uberzeugung von der Unf\"ahigkeit der Feldtheorie, diese Probleme mit ihren Methoden zu 1\"osen, auf Vorurteilen beruht.''}
\end{quote}

\begin{quote}
\textit{``It will be shown that the conviction of the inability of field theory to solve these problems\footnote{The problems refer to the inability of field theory to give account for a molecular structure and to describe quantum phenomena with methods of field theory \cite[p.347]{einstein1936physik}} rests on a prejudice.''}
\end{quote}

and

\begin{quote}
\textit{``Angesichts dieser Sachlage erscheint es mir durchaus gerechtfertigt, die Frage ernsthaft zu erw\"agen, ob nicht doch die Grundlage der Feldphysik mit den Quanten-Tatsachen vereinbar ist.''}
\end{quote}

\begin{quote}
\textit{``In view of this situation it seems appropriate to me to reconsider the question if a unification of the foundations of field theory and quantum facts is possible\footnote{Translation by the author}.''}
\end{quote}

This sums up the scientific program Einstein attempted to present. At the beginning, he refers to a comparison between Quantum Mechanics and statistical mechanics, and states that a statistical or effective theory is not an appropriate starting point to develop a complete\footnote{In the sense of \textit{Born completeness} and \textit{theory completeness}} theory. In his view, Quantum Mechanics was this effective theory that delivers only statistical predictions, but gives no account for the single events. \newline

In a more private environment, a letter from the 16th February, 1936 to his friend Michele Besso, Einstein expresses this point of view explicitly. Since to my knowledge this letter has not been cited in context with the EPR discussion, we want to cite it in full length\footnote{We omit a passage on the political situation in Europe} \cite[p.308]{speziali1972correspondance}.

\begin{quote}
\textit{``Ich halte die statistische Physik trotz all ihrer Erfolge doch f\"ur eine vor\"ubergehende Phase und habe Hoffnung, zu einer wirklich befriedigenden Theorie der Materie zu gelangen. Ich sende Dir gleichzeitig eine kurze Arbeit, die den ersten Schritt darstellt. Das neutrale und das elektrische Teilchen erscheinen gewissermassen als Loch im Raume, derart, dass das metrische Feld in sich selbst zur\"uckkehrt. Der Raum wird als zweischalig dargestellt. In der Schwarzschild´schen strengen zentralsymmetrischen L\"osung erscheint das Teilchen im gew\"ohnlichen Raume als Singularit\"at vom Typus $ 1 - \frac{2m}{r}$. Durch die Substitution $ 1-2m=u^2$ wird das Feld regul\"ar im r-Raume. Wandert $u$ von $-\infty$ bis $+\infty$, so wandert r von $+\infty$ zu $r=2m$ und hierauf wieder zur\"uck zu $r=+\infty$. So kommen beide "Bl\"atter" im Riemann´schen Sinne zustande, die an der "Br\"ucke" $r=2m$  bezw. $u=0$ stetig zusammenh\"angen. Aehnlich bei der Elektrizit\"at. 
Die Aufgabe an der ich mit einem jungen Kollegen (russischer Jude) unabl\"assig schwitze ist die Behandlung des Mehrk\"orperproblems auf dieser Basis. Wir haben aber die ernsthaften Schwierigkeiten des Problems bereits \"uberwunden, sodass sich bald zeigen wird was daran ist. Jedenfalls ist es eine wundervolle mathematische Aufgabe.''}
\end{quote}

\begin{quote}
\textit{`` To me statistical physics despite its success is a transitory phase and I have hope that we arrive at a really satisfying theory of matter. I send you enclosed a short work, that represents a first step. The neutral and the electrical particle appear as a hole in space, of this kind that the metric field returns to itself. Space is represented as two sheets. In the strict spherical symmetric solution of Schwarzschild the particle appears in usual space as singularity of the kind $ 1 - \frac{2m}{r}$. By substitution $ 1-2m=u^2$ the field becomes regular in r-space. If  $u$ goes from $-\infty$ to $+\infty$, r is going from $+\infty$ to $r=2m$ and back again to $r=+\infty$. This represents both "sheets", that are connected by the "bridge" at $r=2m$  respectively $u=0$. Likewise as it is in electricity. 
The challenge a young college (russian jew) and I labour away over is the many-body-problem on that basis. We have conquered the serious problems already, so it will show soon if there is something serious about it. Anyway it is a wonderful mathematical problem.\footnote{Translation by the author}''}
\end{quote}

The connection between Einstein's critique on Quantum Mechanics, specifically its statistical character as well as the problems of the localizability of particles, and the connection to the Einstein-Rosen bridge approach, could not be closer than presented here. \newline

Einstein's unified field theory program failed for reasons we will not discuss in this work. For a comprehensible presentation of this field theoretic program, refer to \cite[]{van2010einstein}. Einstein expressed his hopes how the program would succeed and how it might contain the quantum phenomena \cite[]{einstein1936physik}: 

\begin{quote}
\textit{``Erst die Untersuchung des Mehr-Br\"ucken-Problems kann zeigen, ob diese theoretische Methode eine Erkl\"arung f\"ur die empirisch erwiesene Massengleichheit der Teilchen in der Natur liefert, und ob sie den von der Quantenmechanik so wunderbar erfassten Tatsachen gerecht wird.''}
\end{quote}

\begin{quote}
\textit{``Only the examination of the many-bridge-problem can show if this theoretical model provides an explanation for the empirical proven equality of masses of particles in nature, and if it can reproduce the facts that are represented by Quantum Mechanics in such a delightful way\footnote{Translation by the author}.''}
\end{quote}

These hopes of Einstein seem to be carried on in recent program claims that reconsider the idea of a deep connection between entanglement and Einstein-Rosen bridge solutions.

\subsection{Recent claims}

Recents claims bring back the EPR-ER idea in the context of Quantum Gravity \cite[]{maldacena2013cool}:

\begin{quote}
\textit{``General relativity contains solutions in which two distant black holes are connected through the interior via a wormhole, or Einstein-Rosen bridge. These solutions can be interpreted as maximally entangled states of two black holes that form a complex EPR pair. We suggest that similar bridges might be present for more general entangled states.''}
\end{quote}

and \cite[]{susskind2016copenhagen}:

\begin{quote}
\textit{``Now I feel that our current views of Quantum Mechanics are provisional; it's the best we can do without a much deeper understanding of its connection with gravity, but it's not final. The reason involves a very particular development, the so called ER=EPR principle. ER=EPR tells us that the immensely complicated network of entangled subsystems that comprises the universe is also an immensely complicated (and technically complex) network of Einstein-Rosen bridges.''}
\end{quote}

It is remarkable that the critique by \citeauthor{susskind2016copenhagen} sounds not that different from Einstein's reservations on Quantum Mechanics. The term "provisional" and the statement that Quantum Mechanics is not to be seen as final can be understood as synonymous to Einstein's claims for incompleteness of Quantum Mechanics, in the sense of \textit{theory completeness}. \newline


\newpage
\section{Conclusion}

It is Einstein's merit that he raised questions about the foundations of Quantum Mechanics that are still relevant and bother contemporary physics. In this light, it is more than just history of science to reconsider the old arguments. The EPR-argument is such an argument that is still relevant to the discussions on the interpretation of Quantum Mechanics. If one browses through Einstein's papers of the year 1935, the ER-paper coincides with the EPR-paper, and a closer look reveals that there are further interrelations.  \newline

We presented several notions of completeness that can be identified in the EPR-paper, the ER-paper and its accompanying sources: \textit{Theory completeness}, which refers only to the elements of the theory, \textit{bijective completeness}, or the condition of completeness, as it was named in the EPR-paper, \textit{Born completeness} and \textit{Schr\"odinger completeness}. Recent conceptions of \textit{$\psi$-completeness} and \textit{standard completeness} amend the presentation. We argued that \textit{$\psi$-completeness} is not identical to \textit{bijective completeness} based on the type of elements that both notions of completeness refer to.   \newline 

Einstein's intention of writing the EPR paper was to demonstrate the incompleteness of Quantum Mechanics,\footnote{We identified \textit{Schr\"odinger completeness} as the completeness he challenged in the EPR paper} to justify his ambitions to look for a complete (in the sense of \textit{theory completeness}) theory. This new theory should fulfill \textit{theory completeness} and \textit{Born completeness}, explain the atomistic character of Quantum Mechanics, which turned out to be the singularity-free description of particles, and reproduce the quantum phenomena, which refers to the structure and formalism of Quantum Mechanics. \newline

With the ER-paper, Einstein presented this new theory, which developed into his attempts to present a unified field theory. His critique on Quantum Mechanics lasted in parallel to his ongoing work on the unified field theory, but focused later on a position that was easier to hold. He criticized \textit{Born completeness}, because to him it was undeniable that a new theory is needed, and he shifted the effort from critique on Quantum Mechanics to development of a unified field theory. \newline

The tempting thought to explain nonlocal phenomena inside Quantum Mechanics with ER-bridges is nothing Einstein would have considered, because to him separability was a principle applied to show that Quantum Mechanics is incomplete, in the sense of \textit{Schr\"odinger completeness} independently of a measurement carried out on the system. 

By applying the separation principle and exclusion of the measurement process from the consideration, Einstein aimed to show that the statistical character of Quantum Mechanics is not owed to the measurement process, but resides only in the state vector formalism and the Schr\"odinger evolution. The state vector formalism is fully deterministic, and by that he concludes that the formalism of Quantum Mechanics in incomplete. \newline


In the literature about the EPR-argument, the statistical character of Quantum Mechanics plays a key role in Einstein's critique on Quantum Mechanics, and it truly deserves this role. We emphasized one further essential motivation for Einstein's attempt to reconstruct Quantum Mechanics, the "atomistic structure," or as we identified it: the problem of an absolute localization of particles. In classical field theory, the attempts to give a localization of particles result in field singularities. For Einstein, this problem was an incompleteness in general relativity, in the sense of \textit{theory completeness}. He aimed to solve this problem with a new complete theory, which he first presented in outlines in the ER-paper. The ER-bridge was intended merely to solve the problem of localization of particles; but is also would provide a singularity-free representation of particles and, according to Einstein, would be consistent with Heisenberg's uncertainty principle. It must have been insights like this that encouraged him to hope that this approach promised that quantum phenomena might emerge in his new theory and, by that, reconstruct Quantum Mechanics out of field theory. \newline 

Unfortunately, these hopes of Einstein failed, and until the end of his life, he could not present a satisfying solution to this problem. Though we showed that Einstein never had in mind resolving nonlocality issues in Quantum Mechanics by Einstein-Rosen bridges, because of his non-acceptance of nonlocality as a feasible concept in a theory. \newline

Recent programmatic approaches that claim a EPR=ER principle have been presented. We cannot estimate if the EPR=ER approaches will contribute to the development of a unified theory of quantum gravity, but the idea is promising, and if the program would turn out to be successful, it would be a late gratification for Einstein's field theory program to unify Quantum Mechanics and gravity, even if this success was not in the originally intended direction. It would also underline the extent of Einstein's intuition, which brought forward foundational problems in the interpretation of Quantum Mechanics and probably indicated the right direction in this quest.

\section{Acknowledgments}

I would like to thank the head of the Quantum Particle Workgroup Beatrix Hiesmayr for her support and Basil Hiley, Chris Fuchs, Stefanie Lietze, Agnes Rettelbach, and several other colleagues for fruitful discussions.  For the realisation of the graphical elements and illustrations, I would like to thank Isabella W\"ober.

\newpage

\bibliography{G:/Physik/Bibfiles/Quantenmechanik}

\end{document}